\documentclass[a4paper,fleqn,usenatbib]{mnras}

\usepackage[T1]{fontenc}
\usepackage{ae,aecompl}
\usepackage{amsmath}
\usepackage{pbox}

\usepackage{graphicx}	
\usepackage{amsmath}	
\usepackage{amssymb}	
\usepackage{gensymb}
\usepackage{subfig}

\usepackage{todonotes}

\title[The luminosity dependence of disc winds in LMXBs]{The luminosity dependence of thermally-driven disc winds in low-mass X-ray binaries}

\author[N. Higginbottom et. al]
{Nick Higginbottom,$^{1}$\thanks{E-mail: nick\_higginbottom@fastmail.fm}
Christian Knigge$^{1}$, Knox S. Long$^{2,3}$, 
James H. Matthews$^{4}$ \newauthor{and
Edward J. Parkinson$^{1}$.}
\\
$^{1}$School of Physics and Astronomy, University of Southampton, Highfield, Southampton, SO17 1BJ, UK\\
$^{2}$Space Telescope Science Institute, 3700 San Martin Drive, Baltimore, MD, 21218, USA\\
$^{3}$Eureka Scientific Inc., 2542 Delmar Avenue, Suite 100, Oakland, CA, 94602-3017, USA\\
$^{4}$University of Oxford, Astrophysics, Keble Road, Oxford OX1 3RH, UK\\
$^{5}$School of Mathematics and Physics, Queen's University Belfast, University Road, Belfast 
BT7 1NN, UK\\
}

\date{Accepted XXX. Received YYY; in original form ZZZ}

\pubyear{2018}

\hypersetup{draft}
\begin{document}
\label{firstpage}
\pagerange{\pageref{firstpage}--\pageref{lastpage}}
\maketitle

\begin{abstract}

  We have carried out radiation-hydrodynamic simulations of
  thermally-driven accretion disc winds in low-mass X-ray binaries. Our main
  goal is to study the luminosity dependence of these outflows
  and compare with observations. The 
  simulations span the range 
  $\rm{0.04 \leq L_{acc}/L_{Edd} \leq 1.0}$ and therefore cover most
  of the parameter space in which disc winds have been observed. 
  Using a detailed Monte Carlo treatment of ionization and radiative transfer, 
  we confirm two key results found in earlier simulations that were carried out in the optically thin limit:
  (i) the wind velocity -- and hence the maximum blueshift seen
  in wind-formed absorption lines -- increases with luminosity;
  (ii) the large-scale wind geometry is quasi-spherical, but 
  observable absorption features are preferentially produced along
  high-column equatorial sightlines. In addition, we find that (iii) the wind {\em
  efficiency} always remains approximately
  constant at $\rm{\dot{M}_{wind}/\dot{M}_{acc} \simeq 2}$, a behaviour that is 
  consistent with observations.
  We also present synthetic Fe \textsc{xxv}
  and Fe \textsc{xxvi} absorption line profiles for our simulated disc winds in order
  to illustrate the observational implications of our results.

\end{abstract}

\begin{keywords}
Accretion discs -- hydrodynamics -- methods:numerical -- stars:winds -- X-rays:binaries
\end{keywords}



\section{Introduction}
Signatures of outflowing gas have been observed in essentially all
types of disc-accreting astrophysical systems, from protostars,
cataclysmic variables and X-ray binaries to Seyfert Galaxies and
Quasars. Highly collimated fast jets are often the most spectacular
type of outflow from these systems. However, less collimated, slower
and more highly mass-loaded disc winds are at least as common and can 
actually have a more significant impact, both on the environments of
these systems and on the accretion flow itself. 
 
Low mass X-ray binaries (LMXBs) are systems in which a secondary star transfers mass to a 
compact primary (either a black hole or a neutron star)
via an accretion disc. They are excellent laboratories in which to
study the accretion physics, with lessons learned from LMXBs often finding application also in AGN and other systems \citep{2001MNRAS.323L..26U,2003MNRAS.345L..19M,2004A&A...414..895F,2006MNRAS.372.1366K,
2008Sci...320.1318K,
2012MNRAS.421.2854S,	
2013MNRAS.431.2535S,
2014MNRAS.438.1233S,
2015SciA....1E0686S,
2015MNRAS.448.2430V,
2018MNRAS.481.2140A}.
In particular, LMXBs exhibit dramatic changes in their spectra and luminosity over timescales on the order of days
\cite[e.g.][]{1999ApJ...520..776S,2004ApJ...610..378P}, which can be
linked to changes in the nature of the accretion flow
\cite[e.g.][]{1995PASP..107.1207N,2012Sci...337..540F}. Disc winds are 
seen when systems are in the ``high-soft'' state, during which the accretion disc is thought to extend all the way to the 
central compact object \cite[][although see \citealt{2016ApJ...830L...5H}]{2012MNRAS.422L..11P}. Disc wind signatures are not generally
observed in the ``low-hard'' state, during which the inner disc appears to be truncated, and the X-ray emission is dominated by a
Comptonised corona. This suggests that disc winds might play a role in
regulating -- and perhaps triggering -- state changes in LMXBs. In
fact, for sufficiently high mass-loss rates, disc winds are {\em
  expected} to destabilize a steady accretion flow through the 
disc \citep{1986ApJ...306...90S}.

We are therefore interested in developing a theoretical model for
these disc winds, with an eye to 
predicting their mass-loss rates and hence testing the possibility
that they are indeed responsible for state changes. Unfortunately, even the
basic driving mechanism for these disc winds remains a topic of much
debate and active research. However, broadly speaking, there are three main contenders --
radiation driving, thermal driving, and magneto-hydrodynamic driving. 

The first mechanism, radiation driving, involves the transfer of
momentum from the radiation field to outflowing matter. This transfer
can take place via Compton scattering or via the scattering of (mainly
UV) photons by bound-bound transitions. If the ionisation state of the gas is 
favourable, the latter ``line-driving'' mechanism can
produce a radiation pressure more than 1000$\times$ higher than that
due to Compton scattering \citep{1975ApJ...195..157C,
1995ApJ...454..410G}. In X-ray binaries, 
the observed absorption lines suggest that the outflow is highly ionised \citep[e.g.][]{2009ApJ...701..865K,
2018ApJ...861...26A}, with an ionization parameter $\xi \geq 3$
\citep{2016AN....337..368D}. In such an environment, line-driving is
unlikely to be important. However, the luminosity of some LMXBs can
approach or even exceed the Eddington limit, so radiation driving via
Compton scattering alone may be sufficient to drive -- or at least
affect -- the outflow.

The second mechanism, thermal driving, produces outflows whenever gas
is heated to a temperature at which the thermal velocity exceeds
the local escape velocity. In this situation, mass-loss is
inevitable. This mechanism is particularly attractive in 
X-ray binaries, where the high-energy radiation emitted close to the
accretor can irradiate the outer disc, producing a high-temperature
surface layer in which thermal speeds can exceed the escape velocity. As a rule of thumb, thermal winds might be expected
to arise at or just inside the Compton radius ($\rm{R_{IC}}$) - the radius at which gas at the Compton temperature ($\rm{T_C}$ - 
the temperature at which the Compton heating and cooling rates balance)
for a given source SED corresponds to a thermal velocity in excess of
the escape velocity \citep{1983ApJ...271...70B}. More specifically,
$R_{IC}$ is given by 
\begin{equation}
R_{IC}=\frac{GM_{BH}\mu m_H}{k_BT_C},
\end{equation}
where $M_{BH}$ is the mass of the central object, $\mu$ is the 
mean molecular mass (which we set to 0.6), and the other symbols have the usual meaning. 
For our SED, $T_C=1.4\times10^7$~K, so $R_{IC}=4.82\times10^{11}$~cm - about $4.6\times10^5$ gravitational
radii.

The third mechanism for driving winds from accretion discs is
magnetohydrodynamic in nature. In the presence of a sufficiently
strong, large-scale magnetic field, an outflow can be driven from the
disc by the magnetic pressure gradient or by centrifugal forces,
depending on the geometry of the field. Centrifugal forces can be
dominant when the magnetic field is inclined by at least 30$^\circ$
with respect to the disc axis.  In such an outflow, ionized material
is loaded onto the magnetic field lines and accelerated like a bead on
a wire as it is forced to co-rotate with the field out to the
Alfv\'{e}n radius \citep{1982MNRAS.199..883B,2009Natur.458..481N}. 
Observations of the disc-wind in GRO J1655-40 in a
peculiar `hypersoft' state suggested that the wind in that case arose
well inside $\rm{R_{IC}}$. Since the luminosity of the system was
thought to be well below the Eddington limit at the time, it was
argued that this outflow was likely to be magnetically driven 
\citep[but also see \citealt{2006ApJ...652L.117N,2015MNRAS.451..475U,2016ApJ...823..159S}]
{1992ApJS...80..753S,2006Natur.441..953M,2008ApJ...680.1359M,2009ApJ...701..865K}.

In reality, all three mechanisms are likely at play simultaneously. 
Perhaps changing in relative importance depending on the geometry and
accretion state of the source in question \cite[e.g.][]{
2007A&ARv..15....1D,2014SSRv..183..323F}, or even working against each 
other as seen by \cite{2018MNRAS.481.2628W} who found magnetic fields
partially suppressed thermally driven winds.
Of the three mechanisms,
thermal driving is particularly interesting in X-ray binaries, because
it is almost certain to operate on some level whenever a
sufficiently large disc is subjected to strong X-ray
irradiation. Indeed, this mechanism might not only be important in
X-ray binaries but also in protoplanetary systems
\cite[e.g.][]{2012MNRAS.422.1880O} and AGN \cite[e.g.][]{2018MNRAS.476.4395B}. 

The existence of thermally driven outflows from accretion discs has
been postulated since such discs were themselves 
first considered \citep{1973A&A....24..337S}. The first detailed
theoretical analysis was carried out by
\citet{1983ApJ...271...70B}. This was further developed by
\citet{1986ApJ...306...90S}, who suggested that the associated mass loss
could be sufficient to destabilise the disc. More recently,
hydrodynamic simulations have confirmed the viability of this driving
mechanism and have also shown that the resulting outflows can produce
detectable observable blue-shifted absorption features
\citep{1996ApJ...461..767W,2006ApJ...652L.117N,2010ApJ...719..515L,2015ApJ...807..107H}.  

In principle, thermal driving can drive very high mass-loss
rates. However, in practice, the actual mass-loss rate in X-ray heated
ouflows depends strongly on the heating and cooling rates in the
irradiated gas \citep{2017ApJ...836...42H}. These rates, 
in turn, depend critically not just on the source SED
\citep{2017MNRAS.467.4161D}, but also on the position- and
frequency-dependent attenuation of the
radiation field by material between the source and the wind launching
region. This attenuation represents a non-linear coupling between the
the outflow and the radiation field, which cannot be captured in 
standard hydrodynamic simulations.  

In an effort to improve on this, we have coupled the radiative
transfer code \textsc{python} to the hydrodynamics code
\textsc{zeus}. This allows us to carry out the first {\em
radiation-}hydrodynamic (RHD) simulations of thermally driven disc
winds. In our initial benchmark RHD calculation \citep[hereafter
H18]{2018MNRAS.479.3651H}, attenuation significantly reduced the
thermally-driven mass-loss rate, although the outflow still carried
away mass at more than twice the accretion rate onto the central object.
In addition, reasonable agreement was found between synthetic H- and He-like
 K$\rm{\alpha}$ lines of Fe generated from the simulation and those
seen in \emph{Chandra} observations of the LMXB GRO J1655-40 in the
soft-intermediate state. 
 
For the benchmark RHD simulation presented in H18, we adopted an X-ray
luminosity of 4\% of the Eddington luminosity ($\rm{L_{Edd}}$) for a 7 $M_{\odot}$
central black hole. This very much represents the low end of the
luminosity range in which disc winds have been observed, with the
upper end corresponding to $\simeq L_{Edd}$ \citep[][hereafter
P12]{2012MNRAS.422L..11P}. It is therefore clearly important to ask
whether and how the properties of these outflows depend on the
accretion luminosity. For example, observations suggest that the wind efficiency -- by
which we mean the ratio of the mass-loss rate to the accretion rate
onto the central object -- may increase with luminosity,  
(P12, although this relationship is driven by observations of a single
exceptional source). Conversely, recent theoretical work on
thermally-driven disc winds has found that the wind efficiency should
tend to a roughly constant value \citep[][hereafter D18]{2018MNRAS.473..838D}.

Here, we extend our RHD simulations to higher luminosities in order to
study the impact of this parameter on key outflow properties, such as
the wind mass-loss rate, efficiency, geometry and velocity. 
As discussed in Section \ref{section:method}, we  use the same
technique as in H18, with only slight modifications. We present
our results in Section \ref{section:results}, before making comparisons
to observations and earlier theoretical work in Section
\ref{section:discussion}.

\section{Method }
\label{section:method}

As in H18, we use an operator splitting method to link the hydrodynamic code {\sc zeus} 
\citep[][extended by \citealt{2000ApJ...543..686P}]{1992ApJS...80..753S} to our own ionization and
 radiative transfer code {\sc python}\footnote{More information about and the source code for 
 \textsc{python} can be found at \url{https://github.com/agnwinds/python}.} \citep[][extended by 
  \citealt{2005MNRAS.363..615S}, \citealt{,2013MNRAS.436.1390H} and
 \citealt{2015MNRAS.450.3331M}]{2002ApJ...579..725L}. In brief, the scheme works
 by allowing \textsc{zeus} to perform hydrodynamic simulations for a short period of time
 (typically 1000s measured in the timeframe of the simulation; a few seconds of CPU time) 
 with heating and cooling rates that are scaled versions of the
 approximate analytical formulae suggested by \cite{1994ApJ...435..756B} for optically thin conditions. 
 The end state of each brief simulation
 (density, temperature and velocities) is then passed to \textsc{python}, where a Monte-Carlo
 radiation transport (RT) and ionization calculation is carried out to compute the ionization
 state of the gas, and hence the true heating and cooling rates. The frequency-dependent attenuation of the radiation field
 between the central X-ray source and each cell in the simulation is fully taken into account in this step.
 Our Monte Carlo RT method also accounts for multiple scattering, which can be important whenever the line of sight 
 between a cell and the X-ray source is heavily obscured. The cycle is 
 then completed by \textsc{python} passing back the improved estimates of heating and cooling 
 rates to \textsc{zeus}, where they are used to update the scale factors applied to the optically
 thin heating and cooling rates.
  
 Thermally driven winds are expected to operate only beyond $\simeq 0.1 \rm{R_{IC}}$, which is
 independent of the source luminosity \citep[][hereafter W96]{1996ApJ...461..767W}. We therefore leave the 
 simulation geometry unchanged
from H18 and simply increase the luminosity of the central source. We
adopt the same logarithmic grid as in H18, with 200 radial cells running from $0.05R_{IC} \leq r \leq 20 R_{IC}$
and 100 polar cells from $0\degree \leq \theta \leq 90\degree$. The grid concentrates
resolution towards small radii and large angle (the midplane) 
where we expect the density, velocity and temperature to change most quickly.
In addition, we adopt outflow BCs at the inner and outer radial edges, and a minimum density of $10^{-22}\rm{g~cm^{-3}}$ is imposed throughout the grid.
The physical parameters for the different simulations are 
given in the upper part of Table \ref{table:wind_param}. 

We use a constant density boundary condition at $z = 0$ ($\theta=90\degree$), providing
a mass reservoir for any resulting outflow.
The value of this constant density is important, because it determines
whether the simulation grid captures the entire acceleration zone of
the outflow. The thermal equilibrium curve of gas exposed to the SED used in our simulations 
(an $\rm{L(\nu)\propto \exp{(-h\nu/k_BT_x)}}$ Bremsstrahlung spectrum with $\rm{T_x=5.6\times10^7K}$), 
is composed of a cool stable branch, a hot stable branch, and an unstable intermediate branch. More specifically, X-ray irradiated
material in the disc atmosphere can only achieve thermal equilibrium
at $T \simeq ({\rm a~few}) \times 10^4$~K (where line/recombination
cooling balances photoionization heating) or at $T \simeq 10^{6-7}$~K
(where Compton processes dominate). The intermediate temperature
region is thermally unstable. The cool branch is only available at low 
ionization parameters, $\xi = (4\pi F_{irr}) / n_H$ < $\xi_{cool,max}$,
where $F_{irr}$ is the incident irradiating flux above 13.6eV and $n_H$ is the
hydrogen number density. In order to attain the high temperatures 
necessary to launch a thermally driven winds, material in the disc atmosphere must
therefore reach and exceed $\xi_{cool,max}$. Such gas becomes
thermally unstable and heats up rapidly. In doing so, it expands,
driving the wind.
\footnote{Note that an instability is not strictly required to
    produce a thermally driven outflow. However, thermal driving 
    without an instability tends to produce much slower and
    weaker disk winds \citep{2015ApJ...807..107H}.}

In order to fully capture the acceleration zone of the
  flow -- the region where rapid heating and hence acceleration takes place -- 
we have
to ensure that, at each disc radius, at least some material with $\xi
\leq \xi_{cool,max}$ is included in the grid.
In the optically thin limit,
$F_{irr} = L_x / (4\pi R^2)$, this can be achieved simply by setting the 
mid-plane density such that $n_H = L_x / (\xi_{cool,max} R^2)$. The
mid-plane then corresponds exactly to the bottom of the acceleration
zone.  This is optimal, because it maximises the numerical
resolution in the acceleration zone and avoids the
inclusion of cool, static disk material that our simulations are not
designed to model.

However, the RHD simulations presented here are {\em not} performed in
the optically thin limit. Thus the radiation field incident on a given
cell has not just experienced geometric dilution, but also undergone
wavelength-dependent attenuation by intervening material. Thus
$F_{irr} \neq L_x / (4\pi R^2)$, and the shape of the SED seen by a
given cell can be far from the Bremsstrahlung spectrum emitted by the
central source. As a result, neither the ionization parameter in a 
given cell, nor the cell's thermal equilibrium curve are known {\em a
priori} \citep{2017MNRAS.467.4161D}. 

In order to ensure that we nevertheless capture the entire
acceleration zone in our grid, we exploit the fact that attenuation
always {\em reduces} the strength of radiative heating at depth. Thus,
in the presence of attenuation, the base of the acceleration zone
moves up in the atmosphere, towards lower densities. We therefore
simply adopt the same mid-plane density as in optically thin
calculations, as this guarantees that the acceleration zone will
always lie fully within the simulation domain. The cost of this
conservative strategy is that a small wedge of cold, hydrostatically
supported material will also be included in the grid. However,
the angle subtended by this wedge is small, and we retain acceptable 
resolution in the acceleration zone.

All other aspects of the simulation are identical to that described by
H18, including a truncation of the  mid-plane density boundary
condition at $\rm{R=2 R_{IC}}$.

\begin{table}
\begin{tabular}{p{4.0cm}p{0.55cm}p{0.55cm}p{0.55cm}p{0.55cm}}
\hline 
Luminosity ($\rm{L_{edd}}$) & 0.04 & 0.1 & 0.3 & 0.6 \\ 
\hline 
\hline Physical Parameters & & & \\ \hline
$\rm{M_{BH}~(M_{\odot})}$  & \multicolumn{4}{c}{7}\\
$\rm{T_x~(10^7~}$K)  & \multicolumn{4}{c}{5.6}\\
$\rm{\log(\xi_{cold,max})}$&    \multicolumn{4}{c}{1.35}\\
$\rm{T_{eq}(\xi_{cold,max})~(10^3~K)}$ & \multicolumn{4}{c}{50.7}\\
$\rm{R_{IC}~(10^{11}~cm)}$  & 4.82 & 4.82 & 4.82 & 4.82 \\
$\rm{L_x~(10^{37}}~\rm{ergs~s^{-1}})$&  3.3& 8.25 &  24.75 &49.5\\
$\rm{L_x~(L_{crit})}$ & 0.25 & 0.625 & 1.875 & 3.75 \\
$\dot{\rm{M}}_{\rm{acc}}~(10^{17}~\rm{ergs~s^{-1}})$&  4.42 & 11.1 &  33.2 &66.3\\
$\rm{\rho_0~(10^{-12}~g~cm^{-3})}$  & 16.0 & 40 & 120 & 240 \\
\hline
\multicolumn{4}{l}{Derived wind properties}\\
\hline 
$\rm{V_r(max,\theta>60\degree,~{km~s^{-1}})}$ &  259   & 374 & 533    & 642 \\
$\rm{T(max,\theta>60\degree,r>R_{IC},~10^6K}$ & 2.8 & 3.8 & 5.3 & 6.7\\
$\rm{V_{th}(max,\theta>60\degree,~{km~s^{-1}})}$ & 340 & 396& 467 & 526\\
$\rm{N_H~(70\degree)~(10^{22}}~\rm{cm^{-2})}$ & 2.0 & 4.0 &  8.3 & 13 \\
$\rm{N_H~(80\degree)~(10^{22}}~\rm{cm^{-2})}$  & 4.2 & 8.4 & 16 & 25 \\
Angle for EW(Fe \textsc{xxv})>5eV  & 73\degree & 72\degree & 77\degree & 77\degree\\
Angle for EW(Fe \textsc{xxvi})>5eV & 68\degree & 62\degree & 60\degree & 60\degree  \\
${\dot{\rm{M}}_{\rm{wind}}~(10^{18}~g~s^{-1}})$  &  1.1 & 2.7 &  6.7  & 12.7 \\
${\dot{\rm{M}}_{\rm{wind}}~(\dot{\rm{M}}_{\rm{acc}})}$  & 2.5 & 2.4 & 2.0 & 1.9  \\
${0.5\dot{\rm{M}}\rm{V_r^2}~(10^{32}}~\rm{erg~s^{-1}})$  &  4.2 & 21.3 &   109 & 310  \\

\hline
\end{tabular}
\caption{Parameters adopted in the simulations, along with key properties of the resulting outflows.}
\label{table:wind_param}
\end{table}

\section{Results}
\label{section:results}

\begin{figure*}
\includegraphics[width=\columnwidth]{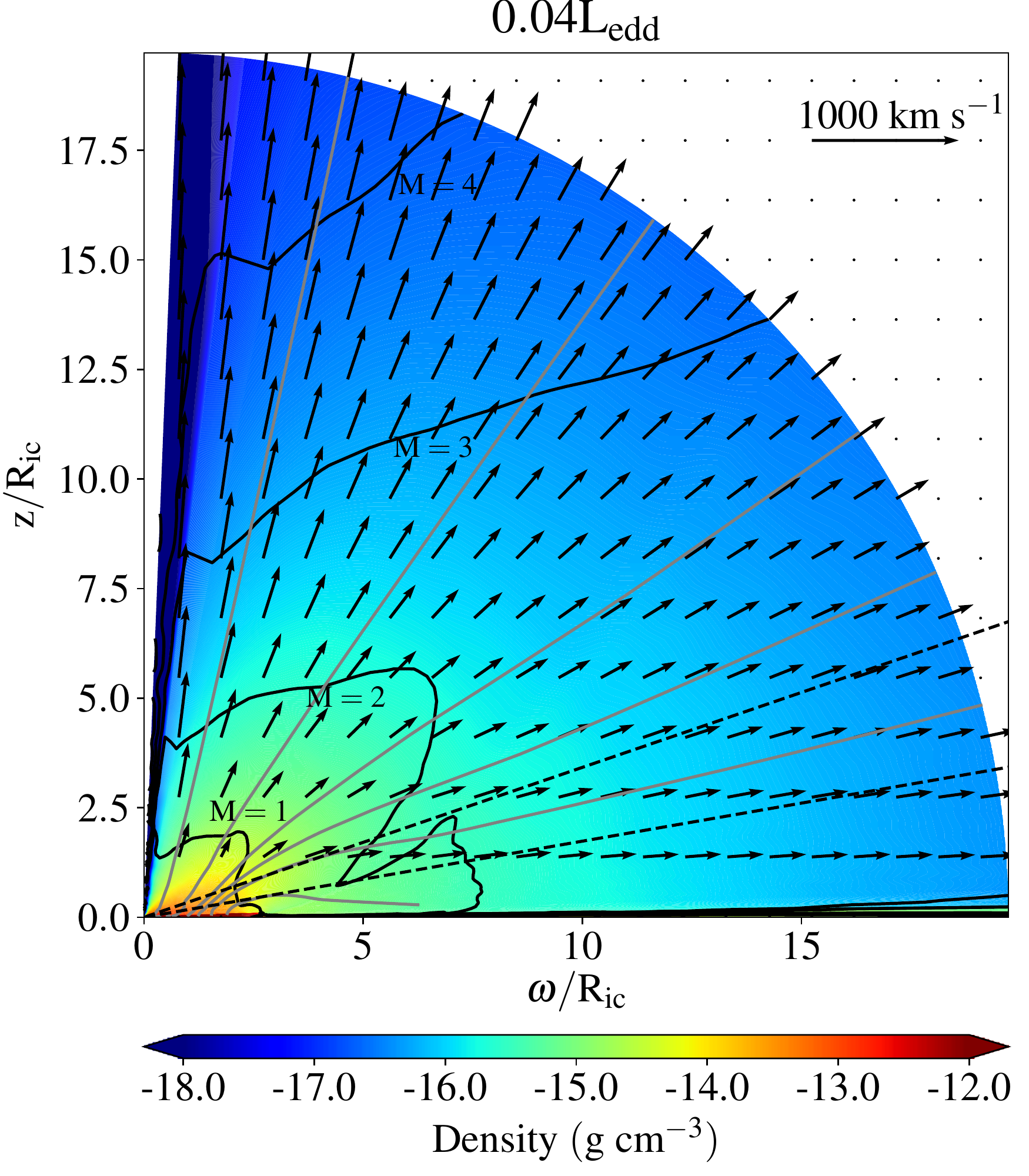}
\includegraphics[width=\columnwidth]{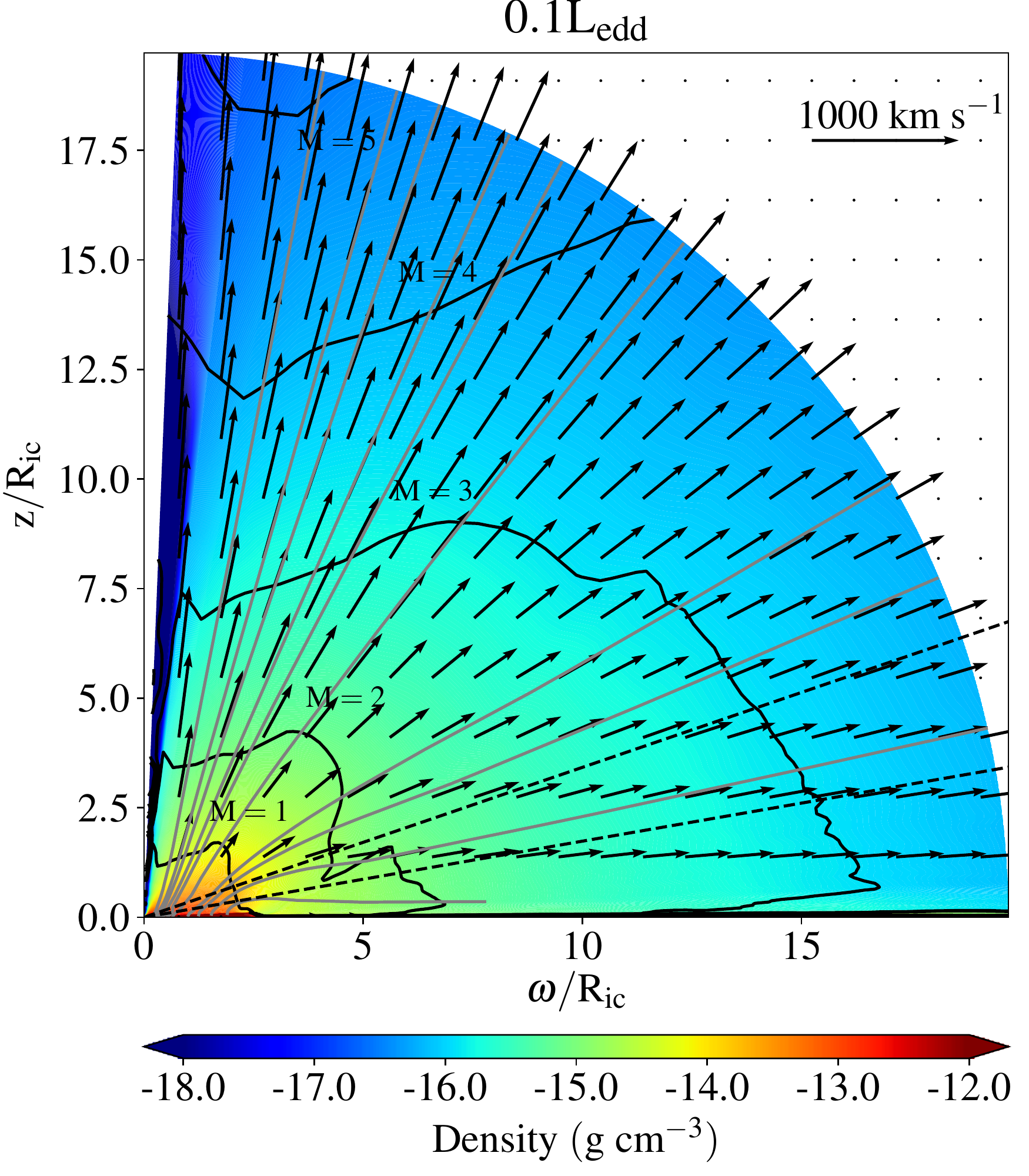}
\includegraphics[width=\columnwidth]{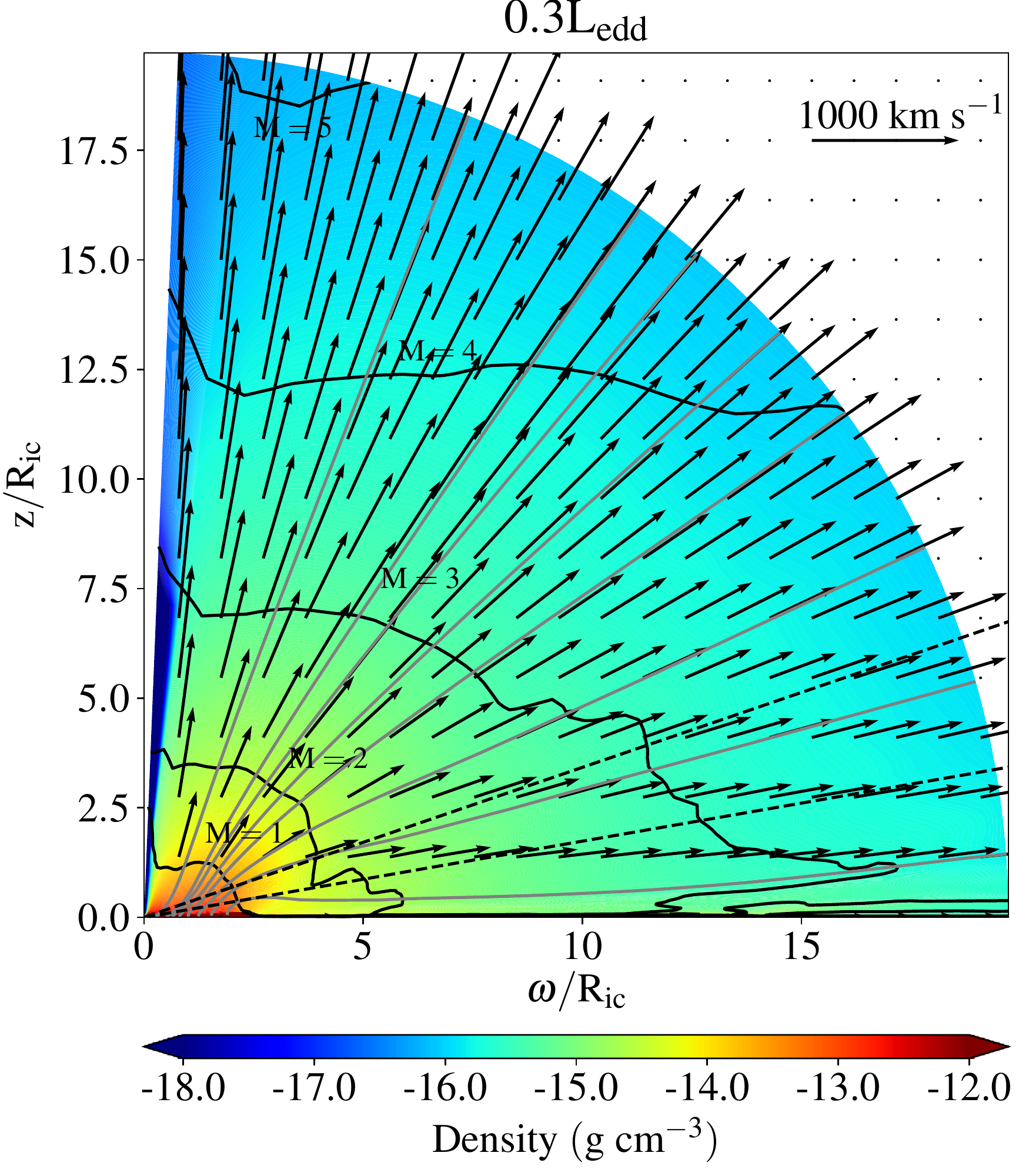}
\includegraphics[width=\columnwidth]{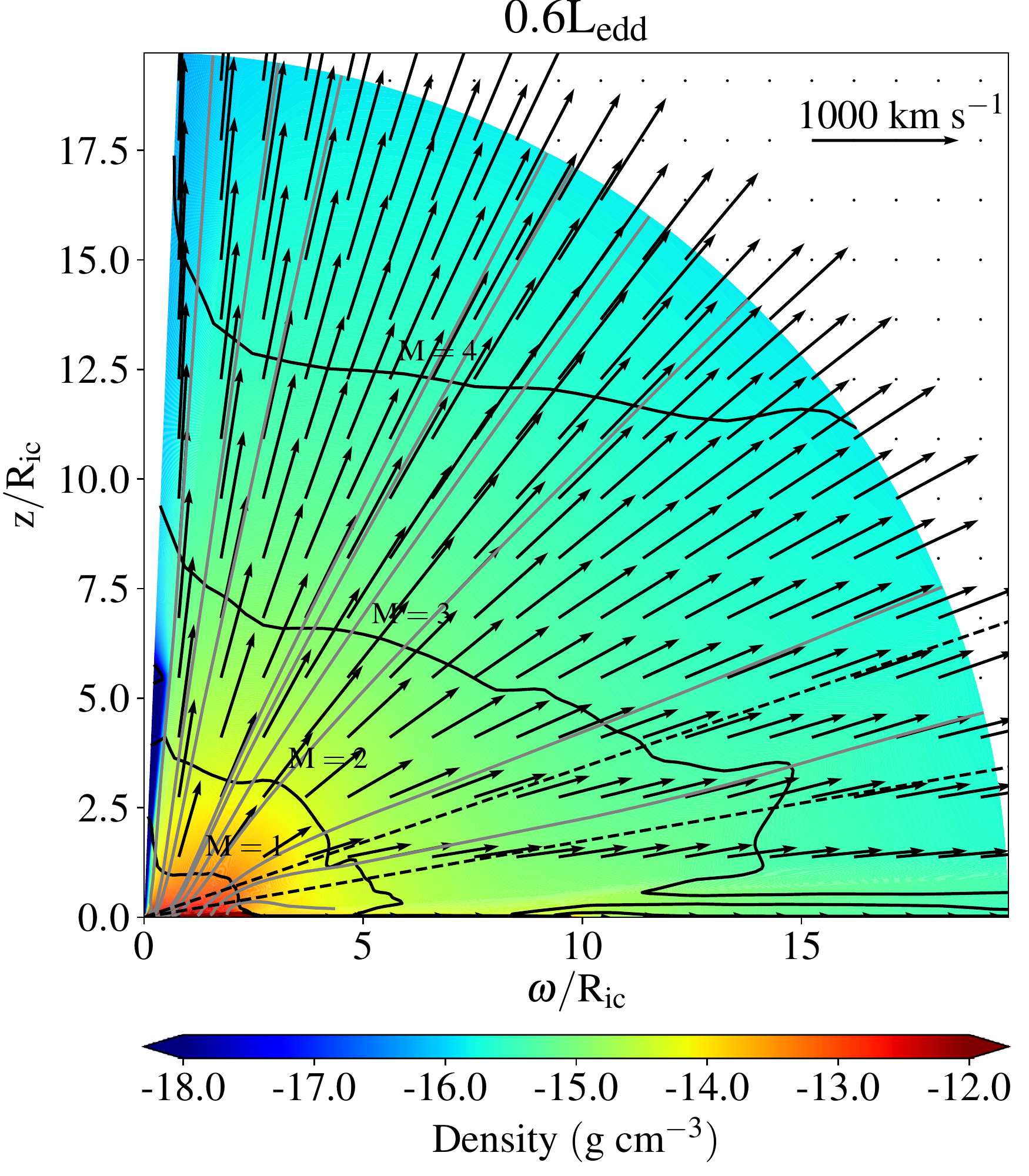}
\caption{The density (colours) and velocity structure (arrows) of the stable final states for 
four different luminosities. Grey lines show streamlines, and the black line shows the location of the Mach surfaces. The two dotted
lines show the location of the $70\degree$ and $80\degree$ sightlines.}
\label{figure:wind_small_image}
\end{figure*}

We have computed new disc-wind models for central source luminosities
in the range $\rm{0.1~L_{Edd} \leq L_x \leq 1.0~L_{Edd}}$ in steps of
$\rm{0.1~L_{Edd}}$. To these, we add our original
$\rm{L_x=0.04~L_{Edd}}$ simulation. 

All of our simulations reach stable states, with static disc-like
wedges forming at the base of the grid.
 As explained in H18, the
reason for this structure is the attenuation of the radiation field by the disc
atmosphere. This attenuation increases with radius, causing the
ionization parameter in the mid-plane of the simulation to drop. As a
result, the vertical height at which $\xi_{cool,max}$ is reached
increases with radius. Since this height marks the effective boundary
between the static disc and the outflow, the net effect is a thin,
slightly convex disc structure. 

Since the mid-plane density in our simulations scales with the central
source luminosity, the column through the disc atmosphere to a given
radius in the mid-plane depends on the luminosity. As a result, the
exact shape of the static disc structure also changes slightly with
luminosity. In all simulations, the disc/wind interface at the
inner edge of the radial grid occurs at around 89\degree~from the
z-axis (i.e. 1\degree~ above the mid-plane). For our $\rm{L_x = 0.04L_{Edd}}$ 
simulation, this angle decreases to about 88\degree~ at the outer
edge; in the $\rm{L_x = L_{Edd}}$ simulation, the disc-wind interface at
the outer edge of the disc lies at an angle of 87\degree.

Above the static disk, at radii within about
   $\rm{0.2~R_{IC}}$, a turbulent clump of subsonic gas forms, 
which can be identified with the corona seen by W96.
As in that work, we find that the size of this hydrostatically supported structure
is roughly independent of luminosity. Further out, the strongly heated gas above 
the midplane disk accelerates and
density contour plots for some of the resulting outflows are shown in
Figure~\ref{figure:wind_small_image}. Some key parameters summarizing
the character of the outflows are given in the lower part of Table~\ref{table:wind_param}. 
Here and elsewhere, we focus mainly on the
results for luminosities $\rm{L \lesssim 0.6 L_{Edd}}$. This is
because our simulations do not include radiation driving, a process
that would become increasingly important as the luminosity approaches $L_{Edd}$. 

Figure~\ref{figure:wind_small_image} clearly shows that the outflow
velocity increases with luminosity, as does the density at a given
point in the wind. As one would expect, the mass-loss rate through the
outer boundary, which is essentially a function of these two
parameters, also increases with luminosity. It is also worth noting
that the density and velocity fields in
Figure~\ref{figure:wind_small_image} do not look particularly 
equatorial. In the following section, we
will consider these critical outflow properties in more detail.

\section{Discussion}
\label{section:discussion}

\subsection{Outflow Geometry}

Wind-formed X-ray absorption lines have so far only been observed in 
systems viewed close to edge-on, and this has been interpreted as
evidence for an {\em equatorial} outflow geometry
(e.g. P12). Yet, as noted above,
Figure~\ref{figure:wind_small_image} does not suggest an equatorial
geometry for our simulated disc winds.

\begin{figure}
\includegraphics[width=\columnwidth]{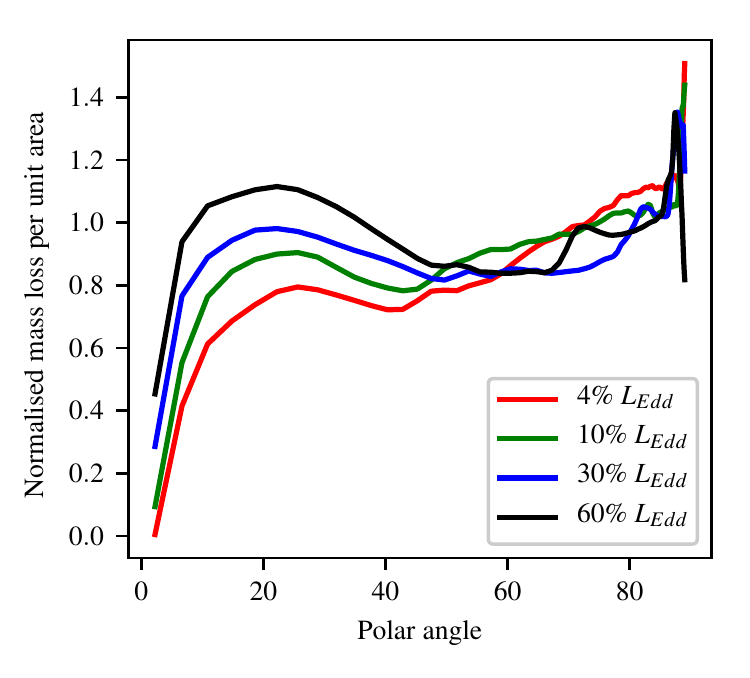}
\caption{The mass-loss rate per unit area through the
outer boundary  as a function 
of polar angle (normalised to the mean mass-loss rate for each
luminosity). Note that, regardless of luminosity, there is no dramatic 
increase in this quantity towards high inclinations. On large scales,
 these outflows are therefore {\em not} strongly equatorial, but rather 
 quasi-spherical.}
\label{figure:geometry1}
\end{figure}

We can quantify the outflow geometry by considering how the mass-loss
rate per unit area depends on polar angle. This is plotted in
Figure~\ref{figure:geometry1}, for the full range of luminosities covered
by our calculations. As suggested by
Figure~\ref{figure:wind_small_image}, all of our thermally driven
winds are essentially quasi-spherical. Except for a narrow $\theta
\lesssim 10$\degree~ `funnel' near the axis, the mass-loss rate per
unit area remains almost constant with polar angle. If the outflows
were strongly equatorial, we would instead expect the mass-loss rate
per unit area to increase strongly towards large polar angles.

This quasi-spherical nature of thermally driven disc winds should
actually come as no surprise. After all, the driving mechanism is
simply thermal expansion, with no intrinsically preferred
direction. On large scales, therefore, {\em any} thermally driven
outflow should be (quasi-)spherical.

\begin{figure*}
\includegraphics[width=\columnwidth]{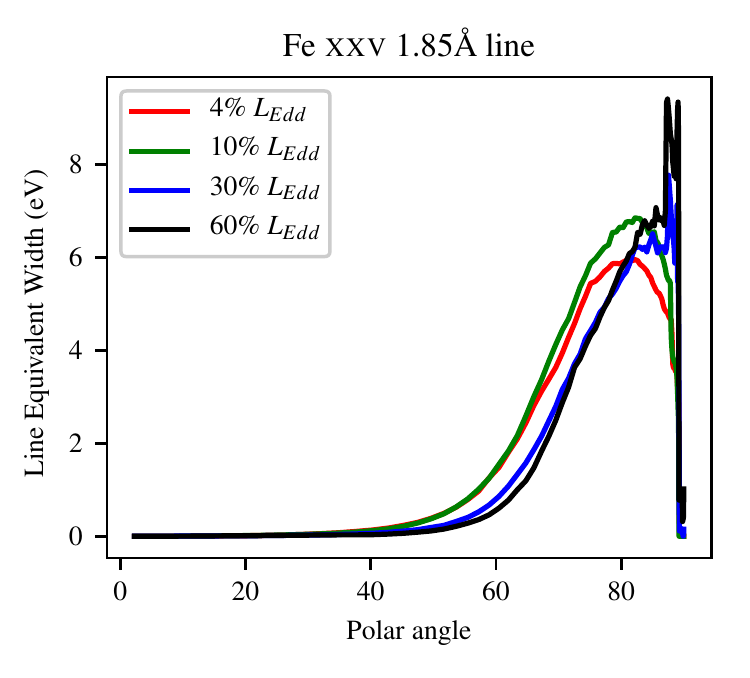}
\includegraphics[width=\columnwidth]{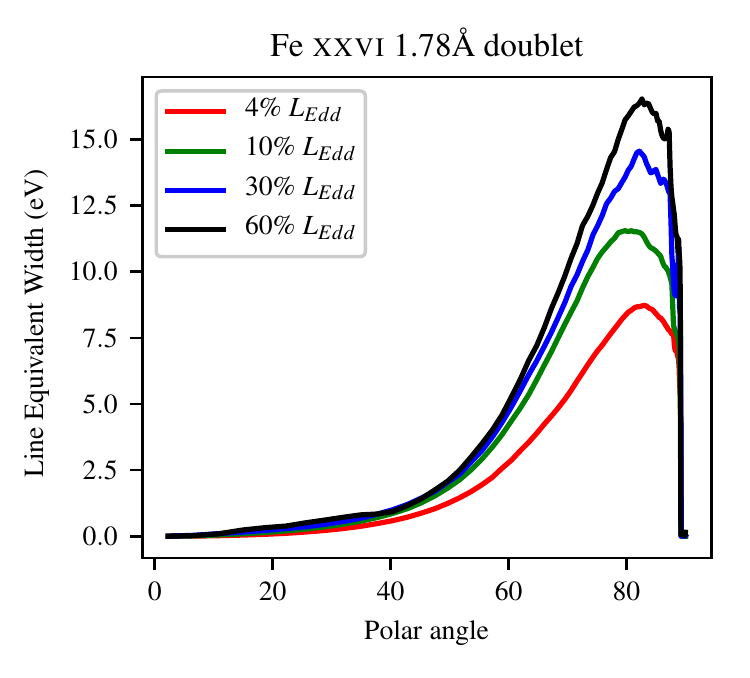}
\caption{The orientation dependence of the wind-formed Fe~\textsc{xxv} 
(1.85~{\AA}) and Fe~\textsc{xxvi} K$\rm{\alpha}$ (1.85~{\AA}) resonance
 lines in our simulations. Even though the {\em mass loss} is quasi-spherical
  (see Figure~\ref{figure:geometry1}), {\em detectable absorption lines} 
  are only expected for equatorial sight lines, in agreement with observations.}
\label{figure:geometry2}
\end{figure*}

So is the geometry of thermally driven disc winds in conflict with
observations? The answer is no. Even though the mass-loss rate
does not vary strongly with polar angle, the column density
through the wind -- and hence the equivalent width of the
wind-formed lines -- does. This is confirmed in
Figure~\ref{figure:geometry2}, which shows the predicted EWs for the
Fe~\textsc{xxv} (1.85~{\AA}) and Fe~\textsc{xxvi} K$\rm{\alpha}$
(1.85~{\AA}) resonance lines (see Section~\ref{lines} for detailed
line profiles for these transitions). In order to be detectable, EWs
$\gtrsim 5$~eV are required, which are only reached for inclination
$i \gtrsim 60^\circ - 70^\circ$ (see Table \ref{table:wind_param} for
more details). This observation is broadly in line with 
the results obtained by W96 in the optically thin limit (c.f. their Figure
29). They, too, found that
the column density of Fe~\textsc{xxv} and Fe~\textsc{xxvi} dramatically 
increased as the inclination angle passed 45\degree.

The reason for this behaviour is simply that the wind emanates from
an accretion disc. For such an outflow, the highest densities will
always be found near the disc plane. As the wind accelerates and
expands away from the disc, its density must drop to maintain mass
continuity. Equatorial sightlines close to the disc plane therefore
encounter more material, even though the outflow itself is not
preferentially focused in this direction.

In addition, W96
found that the column density increases with luminosity. In our
simulations, this behaviour is seen in the Fe~\textsc{xxvi} lines, but
not in the Fe~\textsc{xxv} line. The latter effect is due to
saturation (see Section~\ref{lines}).

\subsection{Mass-loss Rates}

The wind mass-loss rates through the outer boundary
($\dot{\rm{M}}_{\rm{wind}}$) listed in
Table \ref{table:wind_param}   
increase as the luminosity of the central source increases. However,
when we compare these values to the 
accretion rates ($\dot{\rm{M}}_{\rm{acc}}$) onto the central object
implied by our luminosities 
(assuming $\rm{L_x=\eta\dot{M}_{acc}c^2}$ where
the efficiency  $\rm{\eta=0.083}$), we find that
$\dot{\rm{M}}_{\rm{wind}}\simeq2\dot{\rm{M}}_{\rm{acc}}$ for all cases.
\footnote{Note that $\rm{L_{x}}$ and $\dot{\rm{M}}_{\rm{acc}}$
trace the material reaching the innermost part of the disk, whereas
$\dot{\rm{M}}_{\rm{wind}}$ represents the mass-loss rate from the
outer disk. Thus it is not inconsistent to find
$\dot{\rm{M}}_{\rm{wind}} \simeq \dot{\rm{M}}_{\rm{acc}}$, so long as
the secondary can supply $\dot{\rm{M}}_{\rm{2}} \simeq \dot{\rm{M}}_{\rm{wind}} +
\dot{\rm{M}}_{\rm{acc}}$ at the outer disk edge. Once extremely high
mass-loss rates are reached, $\dot{\rm{M}}_{\rm{wind,outer}}/\dot{\rm{M}}_{\rm{acc}} \gtrsim 15$,
the accretion flow may become unstable 
\citep{1986ApJ...306...90S}. However, our solutions are quite far from this
regime.}
This `wind efficiency' is similar to what was found in the
quasi-analytic calculations carried out by D18
 when considering comparable disc sizes and
Compton temperatures. Specifically, they found a peak efficiency of
$\dot{\rm{M}}_{\rm{wind,outer}}\simeq2\dot{\rm{M}}_{\rm{acc}}$ at
luminosities close to the lowest we consider. The wind efficiency then 
decreases slowly with luminosity in their calculations, but always
remains greater than unity. In general, the wind efficiencies in our
RHD simulations are slightly larger than theirs. This is interesting since
we neglect any radiation driving effects, which they include via an
approximate correction.

We have also compared the mass-flux densities in our
simulations against those reported by W96. For the same X-ray
luminosity, we obtain lower mass-loss rates per unit area than found
by them in the optically thin limit (Equation 4.8 in W96). 
We also see a faster drop off with radius, although we have a much reduced radial
dynamic range because our disk is truncated at $\rm{R=2R_{IC}}$. 
Both of these effects are due to a reduction in radiative flux in the acceleration 
zone as a result of attenuation. A drop in the mass-loss rate by about
a factor of 5 -- compared to the optically thin limit -- was also
found in the RHD simulation presented by H18.

\begin{figure}
\includegraphics[width=\columnwidth]{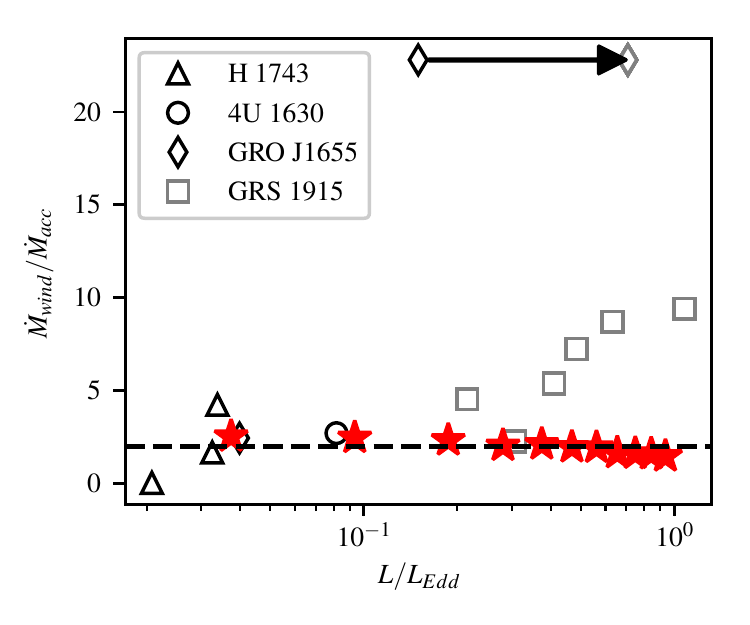}
\caption{$\rm{\dot{M}_{wind}/\dot{M}_{acc}}$ vs. luminosity based upon figure 5 in P12. 
Black symbols are empirical values obtained from {\em Chandra} HETG data for several LMXBs. 
The red stars represent the results of the simulations presented here, and the dotted line
is at $\rm{\dot{M}_{wind}/\dot{M}_{acc}}$=2.0. The luminosity for the highest mass-loss
point for GRO J1655 has ben re-computed by  \protect\cite{2016ApJ...823..159S} and the
arrow and grey diamond show their suggested luminosity of 0.7$\rm{L_{Edd}}$.}
\label{figure:mdot_vs_lum}
\end{figure}

Observationally inferred wind efficiencies have been compiled and
presented by P12, and it is instructive to
compare our results to these. Figure \ref{figure:mdot_vs_lum} shows 
this comparison, with the black symbols referring to the observations
and the red stars showing the results of our simulations. 

At first sight, our prediction of roughly constant wind efficiency
does not appear to agree with the observations. However, the apparent
increase in the observationally inferred wind efficiency with
luminosity is driven entirely by multiple observations of a single
source -- GRS~1915+105 -- at the highest
luminosities, $\rm{L \gtrsim 0.5~L_{Edd}}$. In this regime, radiation
pressure -- which is neglected in our simulations -- is likely to be
important (see above and Section~\ref{section:radiative_driving}). Moreover, even by the
standards of LMXBs, GRS~1915+105 is exceptional. First, it has been accreting
at a significant fraction of the Eddington limit ever since its
discovery in 1992
\citep{1994ApJS...92..469C,2017MNRAS.468.4748C}. Second, it exhibits
extremely unusual spectral states and variability properties
\citep[e.g.][]{2016ApJ...833..165Z}, quite different from the
canonical low/hard and high/soft states seen in most LMXBs. Third, 
with an orbital period of 34 days \citep{2014SSRv..183..223C}, it is
by far the largest known LMXB system and could therefore host a
significantly larger accretion disc than we include in our model.

Given all this, we do not think it is possible to make a meaningful
comparison between our simulations and (at least) the highest
luminosity observations of GRS~1915+105 (hence they are presented
as grey rather than black points in Figures \ref{figure:mdot_vs_lum}
and \ref{figure:ke_vs_lum}). At more moderate luminosities, 
we find reasonable agreement between models and observations over 
roughly an order of magnitude in luminosity,
$\rm{0.02~L_{Edd} \gtrsim L \gtrsim 0.3~L_{Edd}}$. Across this range,
the wind efficiency is consistent with remaining roughly constant at
$\rm{\dot{M}_{wind}/\dot{M}_{acc}} \simeq 2.0$. 

\subsection{X-ray Absorption Lines}
\label{lines}

In order to allow a more direct comparison to observations, we have
also computed synthetic X-ray absorption line profiles for our
simulations. These are calculated using the ray-tracing technique described by
\cite{2017ApJ...836...42H}, using ionic abundances calculated
during the last RT step.

\begin{figure}
\includegraphics[width=\columnwidth]{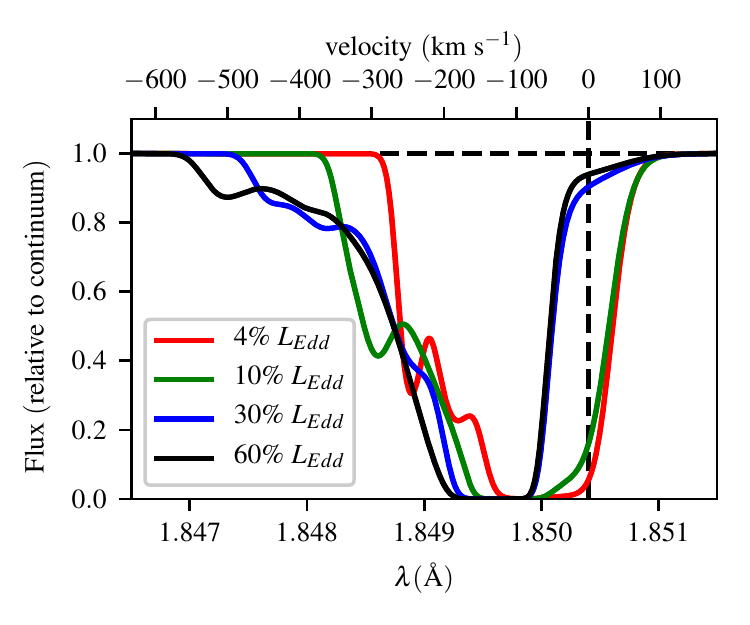}
\caption{Simulated line profile for the Fe~\textsc{xxv} K$\rm{\alpha}$
transition at 1.85~{\AA}, as viewed from $i = 80\degree$ for a range
of luminosities.}
\label{figure:line25}
\end{figure}

In Figure~\ref{figure:line25}, we show the dependence of the
Fe~\textsc{xxv} K$\rm{\alpha}$ transition at 1.85~{\AA} (6.7~keV)
on luminosity, for a representative high-inclination sightline of
80\degree. Since the 
line is always saturated, the overall equivalent width (EW) remains
fairly constant\footnote{We report all our absorption
EWs as positive numbers.}, at about 5-6~eV. However, the blue edge of 
the absorption profiles moves to higher velocities with increasing
luminosity, in line with the faster outflow speeds seen at higher
luminosities in our simulations. 

There is also a notable difference between the line profiles produced
by the two lower-luminosity simulations, and those calculated from the
two higher-luminosity ones. All four models produce significant
absorption at the rest wavelength of the line and at red shifts up to 
about $\rm{v \simeq +100~kms^{-1}}$. This is due to the thermally
broadened line profile associated with stationary or slow-moving
material in the outflow. In the two low-luminosity simulations --
i.e. for $\rm{L \lesssim 0.2~L_{Edd}}$ -- this features is almost black
at $\rm{v \simeq 0~kms^{-1}}$. By contrast, in the two higher
luminosity simulations with $\rm{L \gtrsim 0.2~L_{Edd}}$, saturation 
only sets in around $\rm{v \simeq -100~kms^{-1}}$.

This weakening of zero velocity absorption features is due to the
increased heating rate associated with higher
luminosities. In a thermally driven wind, the velocities in the flow are
expected to be comparable to the characteristic thermal velocity of 
the material in the flow. Stronger heating therefore
produces higher velocities, which is consistent with the disappearance
of low-velocity absorbing material at higher luminosities.

But why is there an apparent step-change in the line profile shape
around $\rm{L \simeq 0.2~L_{Edd}}$? As already noted by
\cite{1983ApJ...271...70B}, one useful way to classify thermally
driven winds is by considering the local heating and dynamical 
time-scales, $t_H$ and $t_{dyn}$, respectively, in the acceleration
zone. If $t_H << t_{dyn}$, gas is heated impulsively and accelerates
quickly. If $t_H >> t_{dyn}$, the outflow is heated steadily as it
rises above the disc and also accelerates more gradually. The critical
luminosity at which $t_H \simeq t_{dyn}$ is expected to be 
\begin{equation}
L_{crit} \simeq 2\times 0.03 T_{C,8}^{-1/2}L_{Edd},
\end{equation}
where $T_{C,8}$ is the Compton temperature in units of $10^8$K,
and the factor of two on the right-hand side is based on the
calculations of W96. Since $T_C=1.4\times10^7$~K
for our adopted SED, $\rm{L_{crit} \simeq 0.16 L_{Edd}}$ in our
simulations. This is close to the luminosity where the zero-velocity
absorption feature disappears in our synthetic line profiles. 
It is therefore tempting to conclude that we may be
seeing a switch from a steadily heated wind to an impulsively heated
one. However, this can only be a speculative conclusion until 
additional simulations are carried out that bracket $\rm{L_{crit}}$
more closely.

\begin{figure}
\includegraphics[width=\columnwidth]{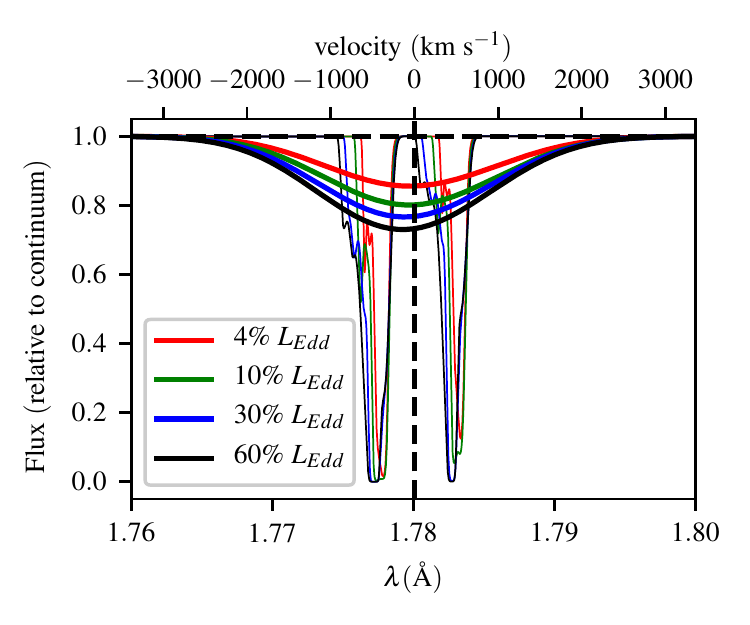}
\caption{Simulated line profile for the Fe~\textsc{xxvi} K$\rm{\alpha}$
doublet, as viewed from $i = 80\degree$ for a range
of luminosities (thin lines) and also smoothed with a Gaussian to represent the appearance when
observed with the \emph{Chandra} HETG with a resolution of 0.012~\AA~(thick lines)}
\label{figure:line26_smooth}
\end{figure}

Another feature that is commonly seen in LMXB winds is the
Fe~\textsc{xxvi} K$\rm{\alpha}$ resonance line at 1.8~{\AA}. This
is actually a doublet, with components at 1.778~{\AA} (6.973~keV) and
1.783~{\AA} (6.952~keV). Figure \ref{figure:line26_smooth} shows the synthetic
absorption lines for this feature. The velocity-dependent
line profile shape of each doublet component is almost identical to
that of the Fe \textsc{xxv} feature in Figure
\ref{figure:line25}. However, since these transitions are slightly less
saturated, they exhibit a larger dynamic range in EW, ranging from
8~eV for $\rm{L = 0.04~L_{Edd}}$ to 16~eV for $\rm{L = 0.6~L_{Edd}}$.
The absorption line obtained from the $\rm{L = L_{Edd}}$ simulation
has an EW of almost 20~eV. However, this should be treated with
caution because of the lack of radiation driving in our simulation.

The current generation of X-ray spectrometers are generally unable
to resolve this doublet. For example, the wavelength resolution of the
first-order spectrum provided by the \emph{Chandra} HETG is
0.012~\AA, compared to the double separation 0.005~\AA\footnote{It is worth noting that
\cite{2015ApJ...814...87M} and \cite{2016ApJ...821L...9M} were able
to extract meaningful {\em third-order} spectra from \emph{Chandra}
observations of a few bright LMXBs. In some of these, the
Fe~\textsc{xxvi} doublet is, in fact, (marginally) resolved.}.
The thick lines in Figure 
\ref{figure:line26_smooth} illustrate the effect of convolving the
synthetic spectrum with a Gaussian representing the first-order \emph{Chandra}
HETG line-spread function.

The presence of detectable Fe~\textsc{xxv} and Fe~\textsc{xxvi}
features in our synthetic spectra is promising. However, do the
properties of these features -- their strength and width/blueshift --
match observations {\em quantitatively}? The observed EWs of
Fe~\textsc{xxv} and Fe~\textsc{xxvi} absorption lines in LMXBs are
typically in the range of 10-30~eV (P12). This is
comparable to the values we measure in our synthetic spectra. However,
 the outflow velocities inferred from
the observed absorption lines are typically 100-3000~$\rm{km~s^{-1}}$
\citep{2016AN....337..368D,2016AN....337..512P}. This range extends to
significantly higher speeds than are found in our simulations.

Part of the reason for this
discrepancy may be continuum driving. Especially for luminosities
approaching ${L_{Edd}}$, this is likely to accelerate material to
higher velocities, but is neglected in our RHD calculations. Higher
wind speeds would also tend to increase the predicted EWs, since they
help to desaturate the line profile by spreading the opacity over a larger
velocity range. Finally, there is some evidence that the observed
absorption lines contain contributions from multiple (low- and
high-velocity) absorption ``zones'' \cite{2015ApJ...814...87M}. If so,
then thermally driven outflows may be associated specifically with 
the low-velocity absorbers.

\subsection{Kinetic Luminosity}

\begin{figure}
\includegraphics[width=\columnwidth]{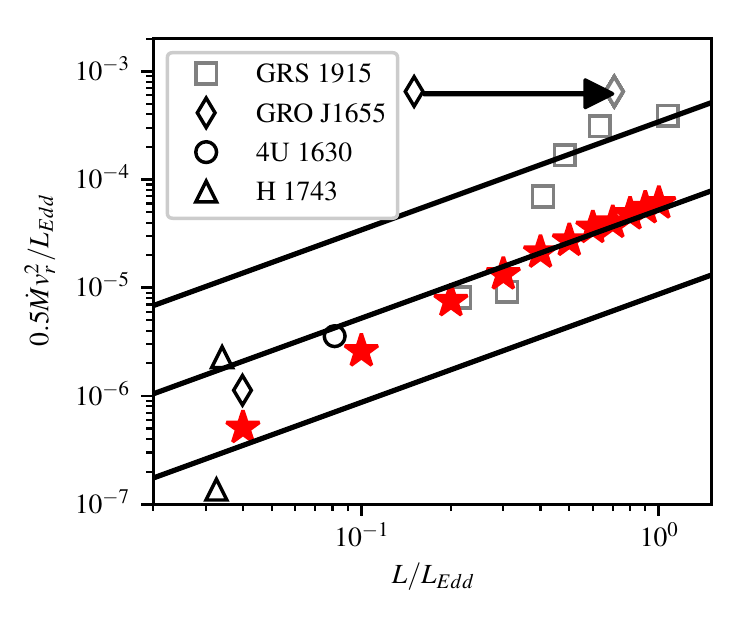}
\caption{Kinetic energy transported by the wind  as a function of source luminosity, both
normalised to the Eddington luminosity. Stars are for this work, the other symbols represent
date from
\protect\cite{2016AN....337..512P}, the symbols code for different sources as in Figure \ref{figure:mdot_vs_lum},
with the arrow and grey diamond showing the \protect\cite{2016ApJ...823..159S} luminosity
value for GRO J1655.
The diagonal lines show theoretical calculations of kinetic luminosity assuming a 0.1 per cent conversion
from luminosity to kinetic energy for discs with 1\degree, 6\degree~ and 40\degree~ total flare.}
\label{figure:ke_vs_lum}
\end{figure}

The kinetic luminosity carried by the disc wind is astrophysically
important for two reasons. First, it represents a potentially
significant non-radiative sink for the available accretion
luminosity. Second, it provides a measure of the likely impact of the
outflow on its surroundings. We therefore estimate the kinetic
luminosity of the disc winds in our RHD simulations by summing
$\frac{1}{2}\dot{M}_i v_{R,i}$ around the outer edge of our
grid. Here, $\dot{M}_i$ and $v_{R,i}$ are the mass-loss rate through,
and radial velocity in, a particular grid-cell $i$ that lies on this
edge. The resulting kinematic luminosities are shown in Figure
\ref{figure:ke_vs_lum} as a function of $L_x$, along with
observationally inferred values taken from
\cite{2016AN....337..512P}, for the same sources shown in
Figure \ref{figure:mdot_vs_lum}.

Fundamentally, thermally driven winds
in LMXBs are driven by the conversion of the X-ray luminosity
intercepted by the accretion disc into the kinetic luminosity of the
outflowing gas. Following \cite{2016AN....337..512P}, we therefore
also include lines corresponding to a
constant efficiency of conversion  in Figure \ref{figure:ke_vs_lum}. These 
assume flared discs with fixed
opening angles of 1\degree,~6\degree and 40\degree.

In our simulations, the hydrostatic ``disc'' near $z =
0$ has an opening angle of $\simeq$4\degree~for low luminosities,
rising to $\simeq$6\degree~for  the $\rm{L=L_{Edd}}$ run. Based on
this, we estimate the conversion efficiency to be 
$\simeq$0.1 per cent at $\rm{L \geq 0.2L_{Edd}}$. In the two lower
luminosity runs, the efficiency is lower, $\simeq$0.04 per cent at
$\rm{0.04~L_{Edd}}$ and $\simeq$0.06 per cent at
$\rm{0.1~L_{Edd}}$. Interestingly, these are again the two runs for
which $\rm{ L \lesssim L_{crit}}$ (c.f. Section~\ref{lines}).

It is worth re-emphasizing here that our RHD simulations are
explicitly designed to {\em not} capture the entire hydrostatic disc
atmosphere. Thus the geometric flaring of the wedge of cool,
quasi-static gas near $z=0$ is not due to the usual convex structure
of hydrostatic $\alpha$-discs. Instead, as discussed
in Section \ref{section:results}, the disc opening angle in our simulations
is driven by the radial dependence of the depth to which X-rays
emitted by the central object can penetrate.

It has been shown that at high Eddington
fractions,  accretion discs tend to puff up to yet larger scale heights
\citep[][but also see
\citealt{2016A&A...587A..13L}]{1988ApJ...332..646A,2005MNRAS.357..295O}. We
may therefore expect real discs -- especially in high-luminosity LMXBs
-- to intercept a greater fraction of the incoming radiation. 
In reality, therefore, the true opening angle might be larger than in
the simulations and this
might explain why the observed kinetic energy flux appears to increase
significantly for $\rm{L \geq 0.3L_{Edd}}$. If the outflows at these
luminosities are thermally driven, with the same efficiency as those
at lower luminosities, we can estimate an implied disk 
opening angle by asking what fraction of the central source luminosity
must be intercepted by the disk surface. This implied disc opening angle is
$\simeq$40\degree. However, in this high luminosity regime,
the efficiency with which radiative luminosity can be
converted to kinetic power may be additionally increased by radiative
driving. This effect would tend to reduce the required opening angle.

\subsection{Radiative Driving}
\label{section:radiative_driving}

At luminosities approaching $\rm{L_{Edd}}$, radiation pressure will
become increasingly important. In general, we expect the plasma above
the disc to by largely ionized, so the momentum transfer between
photons and plasma will be dominated by electron scattering. The net
effect is effectively a reduction in the Compton radius, i.e. the
innermost radius from which a thermal wind can be
launched. Using equations in \cite{2002ApJ...565..455P},
 D18 estimate this reduction to be 
\begin{equation}
\overline{R}_{IC} \rightarrow R_{IC}\left(1-\frac{L}{0.71L_{Edd}}\right),
\end{equation}
which means that the wind can be launched from all radii once 
$\rm{L \geq 0.71L_{Edd}}$.

Since we neglect radiative driving, the mass-loss rates we estimate from
our RHD simulations are, strictly speaking, lower limits for
luminosities approaching or exceeding this value. However, we can make
a rough estimate of the expected size of this effect. In our two
highest luminosity runs ($\rm{L = 0.6L_{Edd}}$ and $\rm{L =
L_{Edd}}$), the mass-loss rate per unit area from the disc is
approximately proportional to $R^{-1.5}$. If all the radii interior to
where the wind current arises were to follow this relationship, the
total mass-loss rate would increase by a factor of about 1.5.  We
might also expect somewhat higher velocities in the outflow, due to
the additional driving force. As noted above, this, in turn, would
then also increase the efficiency with which radiative luminosity is
converted into kinetic luminosity.

\section{Summary}

Thermal driving is an attractive mechanism to explain the outflows
observed in several X-ray binaries seen at high inclinations. We have
previously demonstrated via radiation-hydrodynamic simulations that
the outflow seen in the soft-intermediate state of GRO J1655-40 can be plausibly
modelled as a thermal wind driven by X-ray irradiation. Here we extend
these simulations to higher X-ray luminosities in order to test the
viability of this mechanism more generally. Our main findings are:

\begin{itemize}
\item{The mass-loss rate associated with the thermally driven disc
  winds in our RHD simulations scale roughly linearly with X-ray
  luminosity (and hence accretion rate). Thus the {\em efficiency} of
  these outflows is roughly constant, at
  $\dot{\rm{M}}_{\rm{wind}}/\dot{\rm{M}}_{\rm{acc}} \simeq 2$. This
  agrees with previous theoretical studies and is also in line with
  observations.}
\item{Thermally driven disc winds are {\em not} intrinsically
  equatorial, but rather quasi-spherical. However, since the wind
  densities are highest near the disc plane, the highest {\em columns}
  -- and detectable absorption lines -- are only found for
  high-inclination sightlines, $i \gtrsim 60^\circ - 70^\circ$. This
  is consistent with observations. Thus the absence of wind-formed
  absorption lines from the spectra of low-inclination LMXBs does {\em
    not} require an equatorial outflow geometry.}
\item{The speed of our simulated outflows increases with luminosity, as do the
  blueshifts of the Fe \textsc{xxv} and Fe \textsc{xxvi} absorption
  lines we have calculated for them.}
\item{The kinetic energy carried by the outflow also increases with
  luminosity. The efficiency with which radiative
  luminosity is converted to kinetic luminosity in our simulations is
  $\simeq$0.1 per cent. However, this efficiency depends on both the flare
  angle of the disc and radiation pressure, neither of
  which are self-consistently accounted for in our simulations.}
\end{itemize}
Overall, our results suggest that thermal driving supplemented by 
radiation pressure as $\rm{L\rightarrow L_{Edd}}$ 
(D18) remains 
a good  candidate mechanism for producing the observed disc wind
features in most LMXBs.

\section{acknowledgements}
Calculations in this work made use of the Iridis4 Supercomputer at the
University of Southampton. NSH and CK  acknowledge support by the
Science and Technology Facilities Council grant ST/M001326/1.  
KSL acknowledges the support of NASA for this work through grant 
NNG15PP48P to serve as a 
science adviser to the Astro-H project and JHM is supported by STFC 
grant ST/N000919/1. EJP would like to acknowledge financial support from 
the EPSRC Centre for Doctoral Training in Next Generation Computational 
Modelling grant EP/L015382/1.The authors would
also like to thank the rest of the \textsc{python} team, and to Gabrielle Ponti 
and Chris Done for helpful discussions regarding their papers. 
Lastly, we would like to thank the anonymous referee for a very thorough
review and suggestions for addition comparisons which significantly improved 
the paper.

\bibliographystyle{mnras}
\bibliography{bibliography}()
\label{lastpage}

\bsp	

\end{document}